\documentclass{appolb}
\usepackage{graphicx}

\usepackage[numbers]{natbib}
\usepackage{subcaption}

\begin{document}

\title{Pictures of Particle Production in Proton-Nucleus Collisions}

\author{A. H. Mueller\\
Physics Department, Columbia University, New York, NY 10027
}

\maketitle

\vspace{.2in}
\begin{center}
\large{\it Dedicated to Andrzej Bialas in honor of his 80th birthday.}
\end{center}
\vspace{.3in}

\begin{abstract}
\noindent This work focuses on gluon(jet) production in dilute(proton)-dense(nucleus) collisions. Depending on the frame and gauge, gluon production can be viewed as a freeing of gluons coming from either the proton wave function or from the nucleus wave function. These (apparently) very different pictures must lead to the same result and the purpose of this paper is to see how that happens. The focus is on gluons having $k_\perp\sim Q_S$ or gluons in the scaling region $k_\perp/Q_S\gg 1$. In the McLerran-Venugopalan(MV) model with $k_\perp\sim Q_S$ we are able to derive gluon production in a way that (graphically) manifestly shows $k_\perp$-factorization in terms of the number density of gluons in the nuclear wave function. We presume that this picture, and $k_\perp$-factorization, continues to hold in the presence of small-$x$ evolution although we have not been able to explicitly verify this. Our result is in agreement with usual $k_\perp$-factorization where the gluon number density of the nucleus does not appear in an explicit way.
\end{abstract}
\vspace{.2in}

  
\section{Introduction}

The focus of this paper is particle production in proton-nucleus(pA) collisions at high energies\footnote{For a good review of this topic see chapters 5 and 8 of \cite{1}.}. For explicitness, and for simplicity, we shall often describe this scattering in terms of dipole scattering on a nucleus, but it is always easy to convert our discussion from a dipole to an actual proton, though the formulas will sometimes be a bit more cumbersome. In particular we shall be interested in rapidity regions where the saturation momentum of the proton, if we are in a region where that quantity makes sense, is much less than that of the nucleus. We shall refer to this as the forward(proton) region although the rapidity of the particles being measured may be quite far removed from the rapidity of the proton. Since we are only focused on particle production in rapidity regions where the saturation momentum of the nucleus is significant, what is produced immediately after the collision are (mainly gluon) jets which then will later decay into hadrons. It is these gluon jets which are the center of our study.

The dynamics of how the gluons are produced in a pA collision depends very much on the gauge and frame of reference in which the description is made. Suppose  we choose a frame where the proton is right-moving and the nucleus is left-moving. If one observes produced gluon jets which are right-moving, following the proton, the production process is very simple in $A_+=0$ gauge. In this frame and gauge all gluons in the protons wave function having $k_\perp\ll Q_S$ will be produced as gluon jets of transverse momentum $Q_S$ and with a longitudinal momentum fraction the same as in the wave function\cite{2,3}. ($Q_S$ is the saturation momentum of the nucleus at the rapidity of the produced gluon.) Gluons in the proton are freed mainly through multiple scattering with the dense set of gluons in the nucleus. Because there are in general many scatterings as the gluon from the proton passes through the nucleus, $k_\perp$-factorization is far from manifest\cite{1,4,5,6,7,8,9}. The formula for gluon production, see (\ref{eq:3}) and (\ref{eq:4}) below, is very simple as is the picture of the production. 

Now suppose we shift frame and view the same produced gluons, but now as left-movers following the nucleus. Here the natural gauge to use is $A_-=0$ since the produced gluons can naturally be viewed as coming from the wave function of the left-moving nucleus. However, there are many gluons in the nucleus at the rapidity in question and only a very small fraction of them can be freed by the collision with the proton if (\ref{eq:3}) and (\ref{eq:4}) are to emerge. The main purpose of the present paper is to do this calculation which is presented in Sec.\ref{sec:4}.

Our results agree with $k_\perp$-factorization, however, our picture is more complete in the sense that we can exhibit the Feynman graphs, in Fig.\ref{fig:4}, which can be used to evaluate the production and which naturally have a $k_\perp$-factorization structure.

\section{Two different pictures of forward p-A collisions}
\begin{figure}[htbp]
\centerline{%
\includegraphics[scale=1.0]{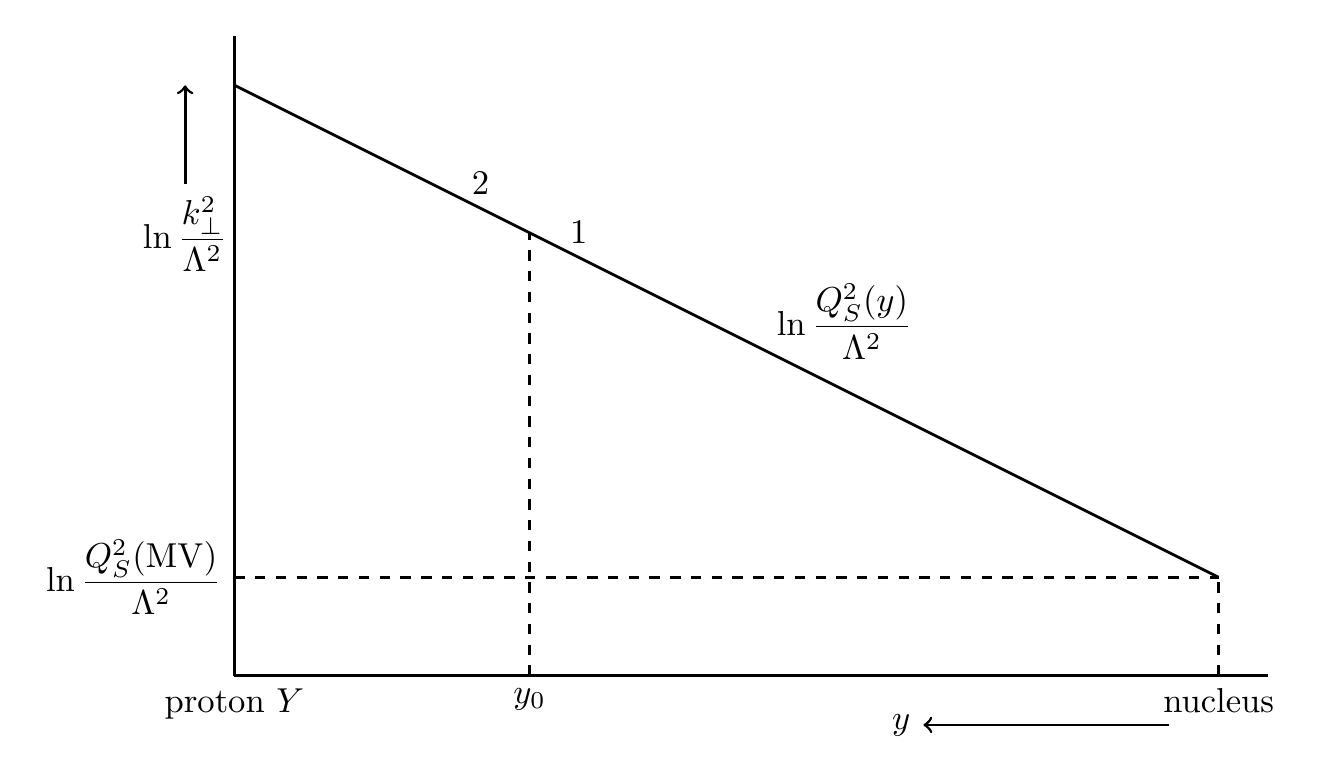}}
\caption{}
\label{fig:1}
\end{figure}

To describe precisely what we are trying to do it is useful to begin with a schematic picture of the saturation momentum, $Q_S^2(y)$, of the nucleus as a function of the rapidity, measured from the rapidity of the nucleus itself. This picture is shown in Fig.\ref{fig:1}. It is also convenient to define three different frames of reference. The A-frame is where the nucleus is at rest and the proton has rapidity $Y$. In the p-frame the proton is at rest and the nucleus has rapidity $Y$. Finally in the $y_0$-frame the nucleus has rapidity $Y-y_0$ and the proton's rapidity is $y_0$. The plot of $Q_S^2$ in Fig.\ref{fig:1} corresponds to the p-frame. In the $y_0$-frame the region of Fig.\ref{fig:1} where $y>y_0$ corresponds to left-moving components of the nucleus' wave function, however $y<y_0$ corresponds to right-movers which are part of the proton's wave function. Thus in the $y_0$-frame only the part of the $Q_S^2(y)$ plot having $y>y_0$ actually corresponds to the saturation scale in the nucleus wave function. In all our discussion where small-$x$ evolution is present we assume a Coulomb gauge is being used where right-moving quanta are effectively in an $A_+=0$ gauge and left-moving gluons in a $A_-=0$ gauge\cite{14}. Thus the nuclear wave function, consisting of left-movers, is described in $A_-=0$ gauge while the proton wave function, consisting of right-movers, is described in $A_+=0$ gauge. In the $y_0$-frame gluons produced at $y_0$ in the figure correspond to gluons having $k_+=k_-=k_\perp/\sqrt{2}$. Gluons produced at points 1 or 2in the figure, $y>y_0$ for point 1 and $y<y_0$ for point 2, are right-moving or left-moving respectively. We call $y_0$ in the forward region of the proton as long as all gluons at $y\geq y_0$ in the proton's wave function are dilute at a transverse momentum scale $k_\perp=Q_S(y_0)$. Finally at $y=0$ in Fig.\ref{fig:1}, and in the horizontal dashed line where evolution is neglected, the saturation momentum, $Q_S(0)$, is given by the McLerran-Venugopalan picture as\cite{11,12,13}
\begin{equation}\label{eq:1}
	Q_S^2(\mathrm{MV})=\frac{4\pi^2\alpha N_C}{N_C^2-1}\rho L x G_p(x,Q_S^2), 
\end{equation} 
where $\rho$ is the density of nucleons in the nucleus, $L$ is the length of nuclear matter at an impact parameter $b$,
\begin{equation}\label{eq:2}
	L=2\sqrt{R^2-b^2}
\end{equation}
and $R$ the nuclear radius. We suppose $xG$ is independent of $x$ in the not too small $x$ region.

Points 1 and 2 in Fig.\ref{fig:1} corresponds to $y$ slightly below $y_0$ and slightly above $y_0$, respectively. In the $y_0$-frame gluons produced at point 2 are right-moving and correspond to gluons from the wave function of the proton which have been freed by the scattering with the nucleus while gluons produced at $y_1$, point 1 in the figure, are left-moving  and come from the nucleus' wave function having been freed by interacting with the the proton. The object of this paper is to see how these two disparate pictures can agree and lead to the same result when $\alpha(y_2-y_1)\ll 1$. In particular we wish to see how gluon production at point 1 can be seen in terms of the gluonic partonic content of the nucleus. As we shall see, calculating gluon production at point 2 in Fig.\ref{fig:1} is very simple in the $y_0$-frame. And this is true whether one neglects evolution, the MV picture, or includes small-$x$ evolution and then evaluates gluon production either in the geometric scaling region, $k_\perp^2/Q_S^2(y_0)\gg 1$, or in the region $k_\perp^2$ on the order of $Q_S^2(y_0)$.

\section{\label{sec:3}Viewing produced particles as coming from the proton}
In this section we evaluate gluon production at $y_2$, in the $y_0$-frame, where the gluon can be viewed as coming from the proton and freed by the nucleus\cite{2,3}. (We do not consider final state emission of gluons from the remnant valence quarks of the proton as such gluons have $k_\perp^2\ll Q_S^2$. In the calculation of \cite{3} such gluons are contained in the first term on the right hand side of Eq.66 of that reference.) Gluons in the proton's wave function will be freed in the collision with the nucleus if those gluons have $k_\perp\leq Q_S(y_0)$ and they will not be freed if $k_\perp\geq Q_S(y_0)$. Therefore, at $y=y_2$
\begin{equation}\label{eq:3}
	\frac{dN}{d^2k_\perp dy}=xG_p(x,Q_S^2(y_0))p(k_\perp,Q_S) 
\end{equation}
where
\begin{equation}\label{eq:4}
	p(k_\perp,Q_S(y_0)=\int\frac{d^2x_\perp}{4\pi^2}e^{-i k_\perp\cdot x_\perp}S(x_\perp,y_0) 
\end{equation}
with $S(x_\perp,y_0)$ the elastic scattering-matrix for a gluon dipole of size $x_\perp$ to scatter on a nucleus at relative rapidity $y_0$. (We suppress the impact parameter dependence in $Q_S$ and in $S$.) In the McLerran-Venugopalan model
\begin{equation}\label{eq:5}
	S=e^{Q_S^2x_\perp^2/4}
\end{equation} 
and, neglecting the logarithmic $x_\perp$-dependence in $Q_S$,
\begin{equation}\label{eq:6}
	p(k_\perp,Q_S)=\frac{e^{-k_\perp^2/Q_S^2}}{\pi Q_S^2}.
\end{equation}
We note that (\ref{eq:6}) can be viewed as a probability distribution of the transverse momenta and that is true of the $p$, defined by (\ref{eq:4}), so long as $S$ is dominantly absorptive.

Eq.(\ref{eq:3}) is very simple. The total number of gluon jets produced is given by $xG_p$ and their distribution is given by $p$. The $x$ in $xG$ is given by
\begin{equation}
	x=k_+/p_+
\end{equation}
where, in the $y_0$-frame, $p_+$ is the light cone momentum of the proton and
\begin{equation}\label{eq:8}
	k_+=e^{(y_2-y_0)}Q_S(y_0)/\sqrt{2}.
\end{equation}
We note that all produced gluons at $y_2$ come from gluons in the proton's wave function which have $k_+$ given by (\ref{eq:8}) (We are neglecting collisional energy loss in the scattering.) however their rapidity in the wave function is given by
\begin{equation}
	y(k_\perp)=y_2+\ln\frac{Q_S(y_0)}{k_\perp}. 
\end{equation}

Independently of the $k_\perp$ that a gluon has in the wave function of the proton, after scattering with the nucleus it ends up with a transverse momentum on the order $Q_S$ thus lowering its rapidity to near $y_2$. ($y_2$ dependences are given in terms of $\alpha y_2$ so that we need not worry about uncertainties in $y$-values on the order of one.)

Eq.(\ref{eq:3}) is correct in a first order hard scattering formalism where the hard scale, $Q_S^2$ in $xG$, can be multiplied by a factor of order one without changing the result significantly and thus our casual statement that all gluons in the proton's wave function having $k_\perp\leq Q_S$ are freed while gluons having $k_\perp\geq Q_S$ are not freed should be understood in this sense. Also, it is relatively easy to see why the elastic $S$-matrix comes into the formula for $p$ in (\ref{eq:4}). To that end we use transverse coordinate space with the produced gluon having $\underline{x}_1$, in the amplitude and $\underline{x}_2$ in the complex conjugate amplitude with $\underline{x}_1-\underline{x}_2=\underline{x}$ in (\ref{eq:4}). But the transverse momentum distribution of gluon going through a nucleus, the $p_\perp$-broadening problem, is given by the $S$-matrix for the elastic scattering of a dipole\cite{14}, $x_\perp$, with the nucleus, thus resulting in (\ref{eq:4}). 

\section{\label{sec:4}Viewing produced particle as coming from the nucleus}
In this section we will rederive (\ref{eq:3}) and (\ref{eq:4}), but now at $y_1$ where the produced gluon(jet) is viewed as left-moving, in the $y_0$-frame, and hence coming from the wave function of the nucleus. (In this section we use the MV model of the nucleus.) Now there are many gluons at $y_1$ in the nucleus' wave function, many more than the number that are freed as given by (\ref{eq:3}). Scattering with a proton frees only a small fraction of the gluons having $k_\perp\sim Q_S(\mathrm{MV})$ in the nucleus' wave function. Our object here is to see how that can come about.

\begin{figure}[htbp]
\centerline{%
\includegraphics[scale=0.6]{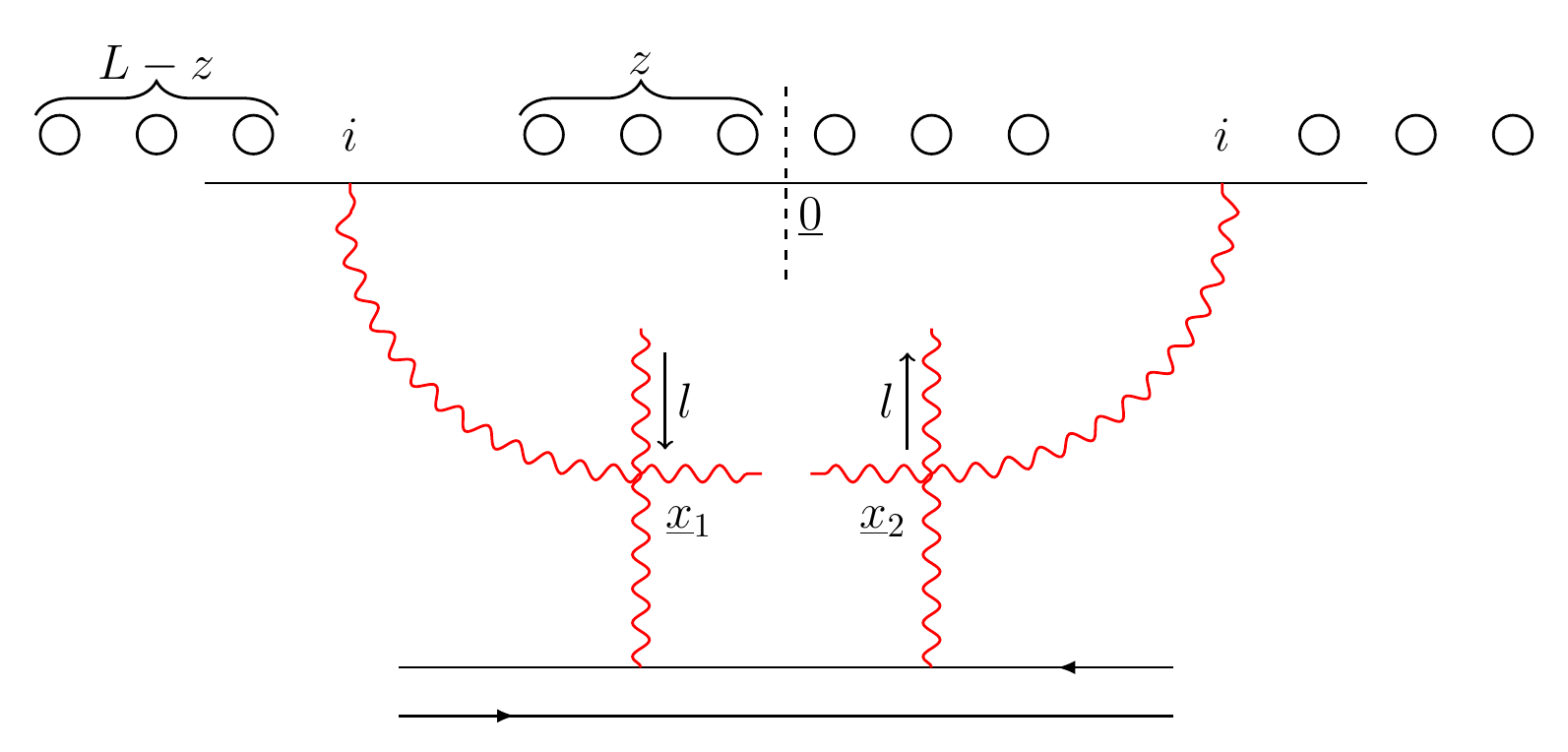}}
\caption{}
\label{fig:2}
\end{figure}

\begin{figure}[htbp]
\centerline{%
\includegraphics[scale=0.6]{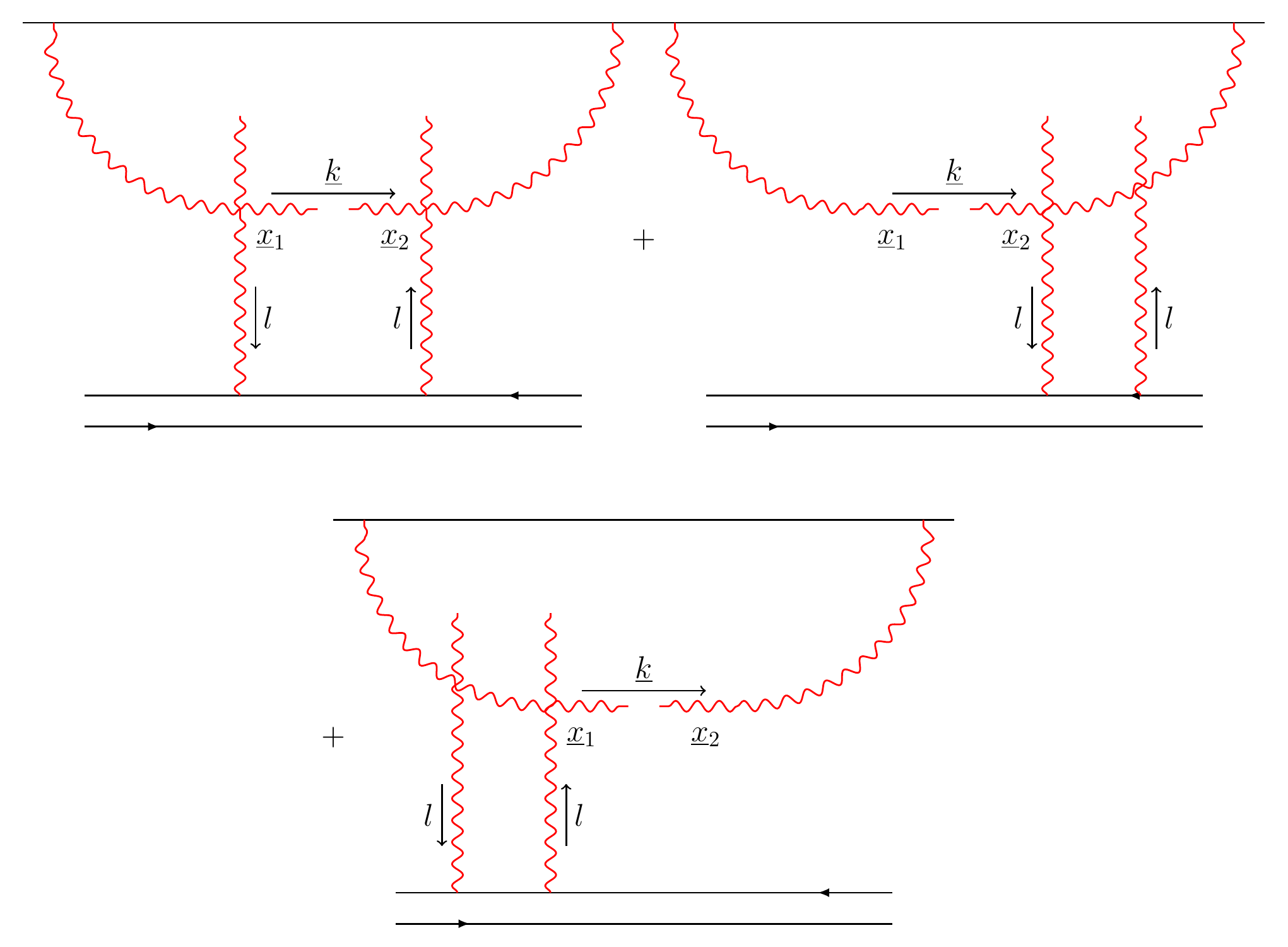}}
\caption{Caption of figure-3}
\label{fig:3}
\end{figure}

\begin{figure}[htbp]
\centering
     \begin{subfigure}{0.48\textwidth}
        \includegraphics[scale=0.6]{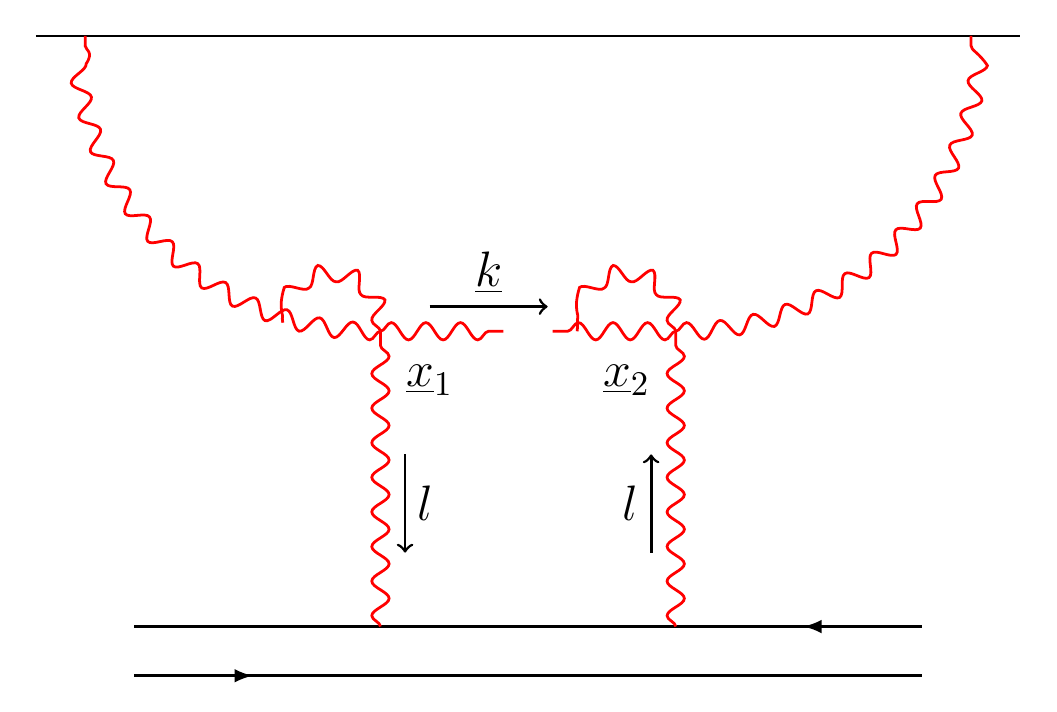}
		  \caption{\label{fig:4a}}
      \end{subfigure}
	  \hfill
      \begin{subfigure}{0.48\textwidth}
         \includegraphics[scale=0.6]{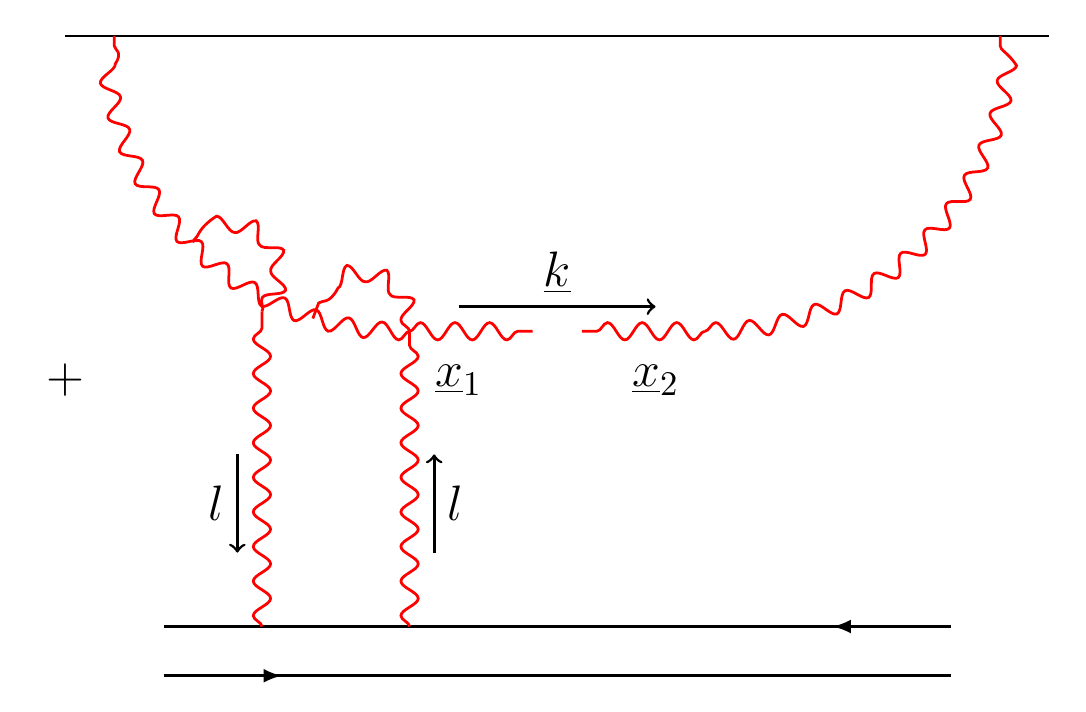}
		   \caption{\label{fig:4b}}
       \end{subfigure}
	   \newline
       \begin{subfigure}{0.48\textwidth}
          \includegraphics[scale=0.6]{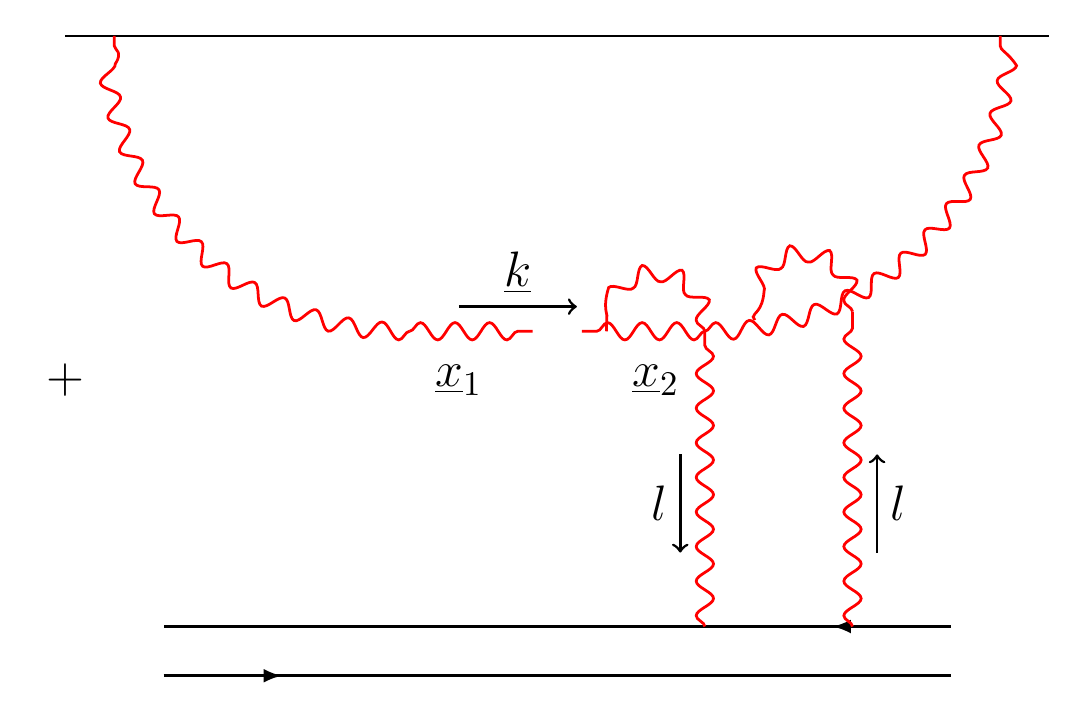}
		  \caption{\label{fig:4c}}
            
        \end{subfigure}
\caption{\label{fig:4}}
\end{figure}

In the MV model a typical contribution to the scattering is shown in Fig.\ref{fig:2} where the produced gluon comes from the $i$-th nucleon in the nucleus lying a distance $z$ from the front of the nucleus. In particular since the produced gluon transverse momentum, $k_\perp$ is the order of $Q_S$ the gluon should come from a specific quark in the $i$-th nucleon as shown in the figure. The scattering with the dipole target then occurs within the color connected part of the active quark in nucleon $i$ and the produced gluon. The key factors are
\begin{equation}\label{eq:10}
	\left(e^{i\underline{l}\cdot\underline{x}_1}-1\right)\left(e^{-i\underline{l}\cdot\underline{x}_2}-1\right)S(\underline{x}_1-\underline{x}_2,z)
\end{equation}
where
\begin{equation}
	S(\underline{x}_1-\underline{x}_2,z)=e^{-\hat{q}(L-z)(\underline{x}_1-\underline{x}_2)^2/4}
\end{equation}
represents the color rotations of the produced gluon by all the nucleons lying behind nucleon $i$. The factors $\left(e^{i\underline{l}\cdot\underline{x}_1}-1\right)$ and $\left(e^{-i\underline{l}\cdot\underline{x}_2}-1\right)$ correspond to exchanged gluons in the amplitude and complex conjugate amplitude, respectively. Now interactions purely in the amplitude or purely in the complex conjugate amplitude do not contribute to gluon production so instead of the first two factors of (\ref{eq:10}) we may instead take (see Fig.\ref{fig:3})
\begin{equation}
	\left(e^{i\underline{l}\cdot\underline{x}_1}-1\right)\left(e^{-i\underline{l}\cdot\underline{x}_2}-1\right)+\frac{1}{2}\left(e^{i\underline{l}\cdot\underline{x}_1}-1\right)^2+\frac{1}{2}\left(e^{-i\underline{l}\cdot\underline{x}_2}-1\right)^2
\end{equation}
but at order $\underline{l}^2$ the expression (\ref{eq:10}) is the same as in Fig.\ref{fig:4}
\begin{equation}\label{eq:expr}
	\left(e^{i\underline{l}\cdot\underline{x}_1-\underline{x}_2}-1\right)
\end{equation}
when an average our direction of $\underline{l}$ is taken. Thus, we use (\ref{eq:expr}) for the scattering with the dipole target and write the gluon production as
\begin{equation}\label{eq:gluon-prod}
	\frac{dN}{d^2kdy}=\int \frac{d^2x}{4\pi^2}e^{-i\underline{k}\cdot\underline{x}}\tilde{N}(\underline{x})2\left(e^{i\underline{l}\cdot\underline{x}-1}\right)\frac{d^2l}{4\pi^2}\frac{g^2C_F}{(\underline{l}^2)^2}\frac{g^2N_C}{N_C^2-1},      
\end{equation}
where we have set $\underline{x}_1-\underline{x}_2$ equal to $\underline{x}$. The factors in (\ref{eq:gluon-prod}) are the following: $\tilde{N}(\underline{x})$ is as in \cite{3} and corresponds to the number of gluons in the wave function of a large nucleus. We shall give $\tilde{N}(\underline{x})$ in a moment. The factor of 2 corresponds to the $l$-line hooking to one or the other of the two parts of the target dipole. The $\left(e^{i\underline{l}\cdot\underline{x}-1}\right)$ is from (\ref{eq:expr}) and gives the phases of the hookings of the $l$-lines to gluons at $\underline{0}$ and $\underline{x}$. $\frac{g^2C_F}{(\underline{l}^2)^2}$ gives the coupling of the $l$-lines to a line of the dipole as well as the $l$-line propagators. Finally $\frac{g^2N_C}{N_C^2-1}$ is the coupling of the $l$-lines to the $k$-lines where we have included a factor of two for a second color connected component in addition the graphs of Fig.\ref{fig:4}. This second color connected component involves final state emission.

In the logarithmic approximation one expands $\left(e^{i\underline{l}\cdot\underline{x}}-1\right)$ to quadractic order in $\underline{l}$. This gives
\begin{equation}\label{eq:13}
	\frac{dN}{d^2kdy}=-xG_D\int d^2\underline{x}\frac{\alpha N_C}{4(N_C^2-1)}\underline{x}^2\tilde{N}(\underline{x})e^{-i\underline{k}\cdot \underline{x}} 
\end{equation}
where
\begin{equation}
	xG_D=\frac{2\alpha C_F}{\pi}\int_{1/x_{01}^2}^{1/\underline{x}^2}\frac{d\underline{l}^2}{\underline{l}^2}=\frac{2\alpha C_F}{\pi}\ln \frac{x_{01}^2}{\underline{x}^2}    
\end{equation}
with $x_{01}$ the target dipole size. From \cite{3,12} the Weizs\"{a}cher-Williams gluon distribution is
\begin{equation}\label{eq:15}
	\tilde{N}(\underline{x})=\frac{N_C^2-1}{\pi^2\alpha N_C\underline{x}^2}\left(1-S(\underline{x})\right) 
\end{equation}
with $S(\underline{x})$ exactly as in (\ref{eq:4}). If $\underline{k}\neq 0$ the final answer is
\begin{equation}\label{eq:16}
	\frac{dN}{d^2kdy}=xG_p(x,Q_A^2)\int \frac{d^2x}{4\pi^2}e^{-i\underline{x}\cdot\underline{x}}S(\underline{x}) 
\end{equation} 
exactly as in Sec.\ref{sec:3}. In going from (\ref{eq:13}) to (\ref{eq:16}) we have replaced the dipole target by a proton target by changing $G_D$ to $G_p$. Eq.(\ref{eq:13}) expresses naturally gluon production, on the scale of $Q_S$, in a $k_\perp$-factorized way in terms of $\tilde{N}$, the Weizs\"{a}cher-Williams gluon distribution in the MV model.

There is an intuitive picture of why the graphs in Fig.\ref{fig:4} might be expected to give (\ref{eq:gluon-prod}), (\ref{eq:13}) and (\ref{eq:16}). View the graphs in terms of a Boltzmann precess where a gluon of momentum $k$ is produced, from a flux of incoming gluons of the nuclear wave function as a gain term and a loss term. Fig.\ref{fig:4a} corresponds to a gain term where $k-l\rightarrow k$ and Figs.\ref{fig:4b} and \ref{fig:4c} correspond to loss terms of gluons having momentum $k$. Gluons that do not interact with the target dipole cannot be freed and are not included in (\ref{eq:gluon-prod}).

The discussion we have given here is based on the McLerran-Venugopalan model. One might hope that (\ref{eq:gluon-prod}) and (\ref{eq:15}) are more general than the MV model and it is a challenge for the future to generalize the current analysis to include small-$x$ evolution.

\nocite{*}

\bibliography{ref-Mueller}{}
\bibliographystyle{apsrev}

\end{document}